# SensIs - Underwater acoustic network for ice-monitoring

Tor Arne Reinen[1], Arne Lie[1], Finn Tore Knudsen[2]

[1] SINTEF ICT
[2] Kongsberg Maritime
Contact email: Tor.A.Reinen@sintef.no

## Abstract

Routing for low latency underwater acoustic network-communication is investigated. The application is monitoring of ice-threats to offshore operations in the Arctic - to provide warnings that enable operators to react to such threats. The scenario produces relatively high traffic load, and the network should favour low delay and adequate reliability rather than energy usage minimization. The ICRP (Information-Carrying based Routing Protocol), originally proposed by Wei Liang et al. [1] in 2007, is chosen as basis. ICRP obtains unicast routing paths by sending data payload as broadcast packets when no route information is available. Thus, data can be delivered without the cost of reactive signalling latency. In this paper we explore the capabilities of a slightly enhanced/adapted ICRP, tailored to the ice monitoring application. By simulations and experiments at sea it is demonstrated that the protocol performs well and can manage the applications high traffic load – this provided that the point-to-point links provide sufficient bit rates and capacity headroom.

## 1  Introduction

The *SensIs* project has developed technology for monitoring of ice-threats to offshore operations in the Arctic. The aim is to provide warnings to enable operators to react to such threats. The target area has been the northern Barents Sea, where the threats are mainly large sea-ice features – thick pack ice and ridges. The project has developed both new acoustic Doppler-based sensors and underwater acoustic network technology for communicating the sensor data. In the following the latter is considered: The design and performance evaluation of a suitable underwater acoustic network. The focus is put on the network routing protocol – how to route sensor output data from node to node through the network, to a common sink. For other protocol components quite conventional choices have been made. For point-to-point connections existing modems have been assumed, but the project has developed application-specific modem-transducers, the impact of which on system performance is also demonstrated. Details of the protocol design, beyond those presented here, can be found in [2]. The SensIs consortium comprises Nortek (ice and current sensors), Kongsberg Maritime (acoustic communication), SINTEF ICT (research) and Statoil (problem owner).

     Each sensor is a bottom-moored upward-looking ultrasound pulse-echo instrument. It measures ice thickness and movement, together with the sea current profile. The instrument is connected to an acoustic modem that provides wireless network capabilities. To monitor ice movement in any direction, the ice sensors should constitute

a circle around the information sink (the operations control centre). A typical topology is shown in Figure 1.

The paper is organized as follows: Section 2 gives an overview of the system design, with emphasis on the routing protocol. Section 3 gives network simulation results, while Section 4 provides corresponding results from the final SensIs sea trials. Finally, Sections 5 and 6 provide a discussion and conclusion of the work.

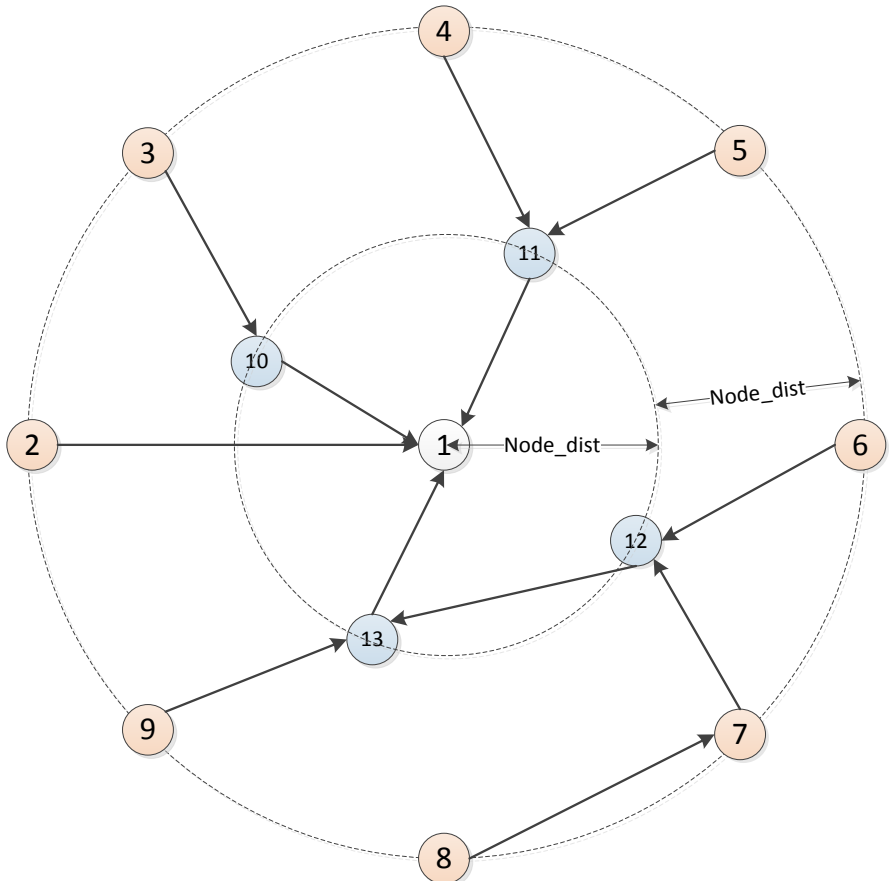

Figure 1 Network topology: Outer circle is sensor nodes, inner circle communication relays and the centre is the information sink. Arrows are examples of communication paths that might be selected by a routing protocol.

## 2  System design

**Topology:** An outer circle comprising 8 sensor nodes and an inner one with 4 communication relay nodes was selected as shown in Figure 1. The node distance parameter in the figure was set to 600 m, so that the outer sensor circle would be 1200 m from the sink. Extreme ice drift velocities in the Barents sea are estimated to be 4 km/h [3]. The 1200 m distance would then correspond to a response time of 18 minutes.

**Sensors:** Nortek Doppler sensors Signature55 and Signature250 were used.

**Modems:** Kongsberg Maritime cNode modems were used, each with an embedded PC added for implementing the communication protocols. For point-to-point communication, three transport formats (TF) were available:
- Direc-Sequence Spread Spectrum (DSSS) at of 200 bits/sec (bps) payload
- DSSS at 400 bps payload

- A channel equalization (turbo) scheme at 1700 bps.

Fragmentation of long messages is available to avoid long-duration packets at low rates.

**Protocols:** The main requirements and challenges are:
- High traffic load. Sensor measurement every 6 seconds.
- Two operational modes
    - Normal mode: Send only each 7th message, i.e. 42 sec intervals.
    - Alarm mode: Send every 3rd message, i.e. 18 seconds intervals.
- Time-variable water channel as is common in underwater acoustics.
- Marginal time for retransmissions due to the traffic load and modem rates.
- Marginal time for routing path selection/optimization correspondingly.

Hence, the protocol system should
- tolerate congestion.
- not impose reactive signalling delay.
- support fast channel dynamics.

To meet these challenges Medium Access (MAC) and routing protocols were chosen as follows:

*MAC protocol*: Carrier Sense Multiple Access (CSMA) Aloha was selected, modified to reply with MAC acknowledgement (ACK) and Automatic repeat request (ARQ) only in cases of data unicast (UC).

*Routing protocol*: This is central to obtaining efficient and low latency communication. It needs to be adaptive due to time-varying channel conditions. Adaptivity must, however, be obtained without time-consuming reactive signalling that would aim to find optimal routes. In addition, we did not want the system to be dependent on node position information.

Through literature search the "Information-carrying based routing protocol" ICRP [1] was found to be promising and chosen as basis for our system. It is suited for the source- to-sink configuration and obtains low latency under varying link qualities.

In brief, ICRP works as follows: At first no preferred route is available, and information is sent as broadcast (BC). Nodes that receive the packet forward it only once. When the packet reaches the sink, a routing-level ACK is returned. This is denoted STATUS, and contains information about the best path. Based on the STATUS message, UC is used for the subsequent transmissions, but the protocol reverts to BC if unicast is unsuccessful.

ICRP was adapted/enhanced for our application (see [2] for details). The main enhancements are:
- Limiting the number of hops per message, in order to reduce the amount of unnecessary message replication in BC.
- Adding patience in UC mode: Permitting a few packets lost (leading to retransmission) before reverting to BC.
- Adding signal to interference and noise ratio (SINR) as part of best-path metric. This information is available from the modems.
- Adding adaptation of modem communication rate to the success-rate of UC communication, using the before-mentioned transport formats.

The two latter points may be considered cross-layer adaptations.

## 3 Protocol simulation

Protocol tuning and adaptation was carried out using DESERT, [4] which is an extension to ns-2/ns-miracle [5] and can be used both for pure simulation and with hardware in the loop. Sound propagation was simulated in a simplistic manner that is available in DESERT, and is known in the acoustic communications community as the "Urick model" or "practical loss model" [6]. This amounts to a simple transmission loss given in dB by $15\log_{10}$ of the distance to the receiver, plus frequency dependent absorption. The deviation from the free space transmission loss of $20\log_{10}$ of distance can be motivated by the assumption that the receiver be able to exploit multipath-energy.

The scenario simulated is that in Figure 1, with 600 m node-distance, as mentioned in section 2. The transmit-power, noise level and modem parameters were combined such that maximum transmission range is just sufficient for the outer nodes to reach the sink in a single hop. Still, multi-hop configurations could well be selected by ICRP due to packet collisions. Constant bitrates and packet intervals were used, and the throughput was measured as function of packet interval length for the three transport formats (TFs) of the modem. Remember from section 2 that normal and alarm modes correspond to 42 and 18 seconds packet interval respectively. The resulting packet delivery ratio (PDR) is shown in percent in Figure 2 for the three TFs. The figure also shows the percentage ICRP STATUS messages. These messages are sent from the sink in reply to BC transmission, and hence indicate that stable UC routes are *not* established. Table 1shows numerical results corresponding to Figure 2, for the normal- and alarm-states.

Table 1 PDR in % at normal and alarm states

| bits/s | Normal | Alarm |
|---|---|---|
| TF1: 200 | 38 | 10 |
| TF2: 400 | 80 | 20 |
| TF3: 1700 | 98 | 88 |

The results clearly demonstrate that the network performance benefits from being able to run at higher bit rates. In particular, TF1 is clearly inadequate, and TF2 is also not satisfactory in alarm mode[1]. TF3 is deemed sufficient according to our application scenario.

Simulations have also been carried out with two real sensors in the loop. This setup gives somewhat less regular traffic input and requires much longer simulation times. The latter, then, practically leads to results with lower statistical confidence. Still the results are entirely in line with the above, essentially confirming that only TF3 gives fully sufficient performance for the target application.

---

[1] whether or not many (likely similar) messages could be disregarded in alarm mode depends on the high-level usage of the information. But this is outside of our scope.

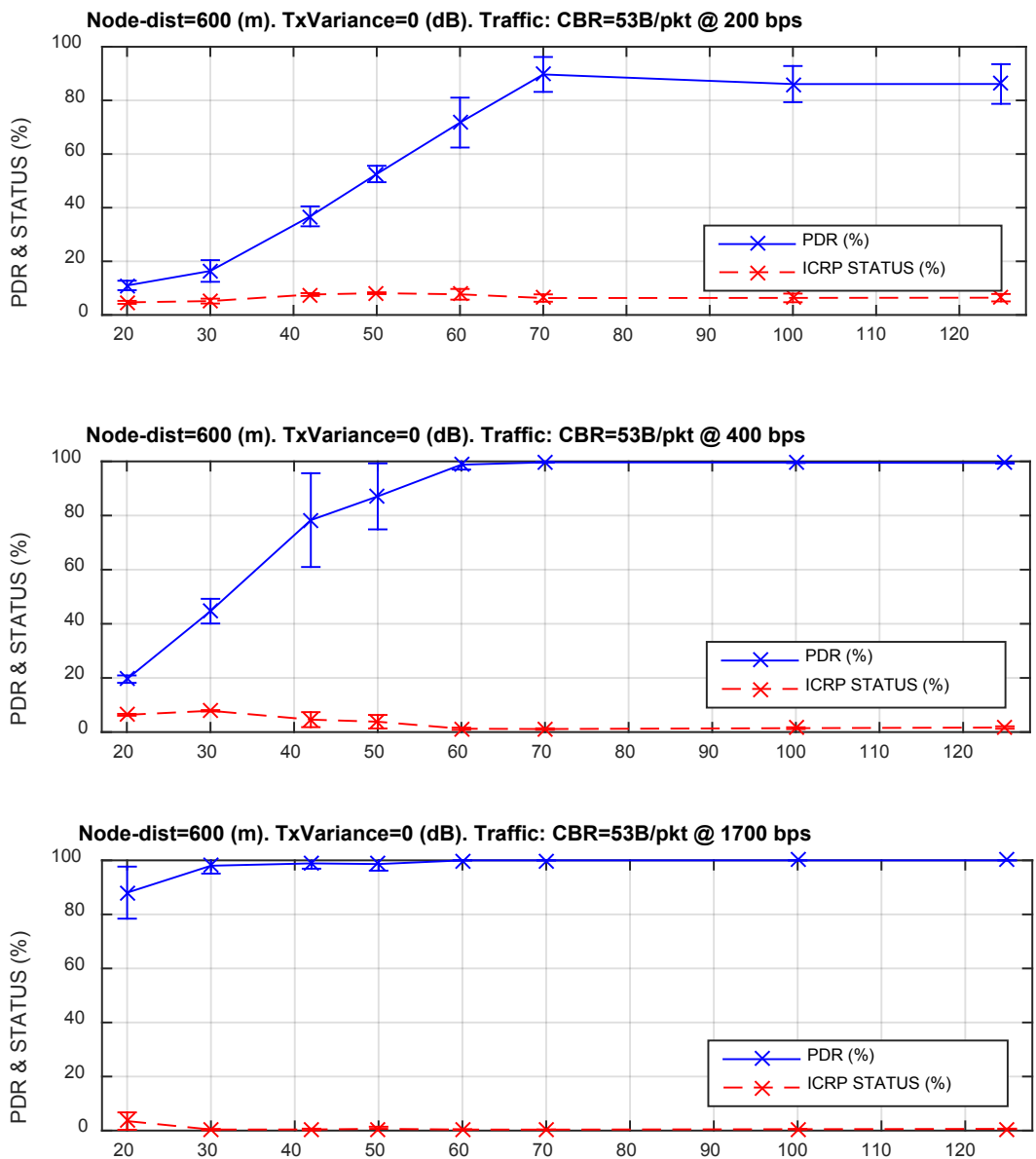

Figure 2 Traffic simulation results: % PDR and % STATUS messages vs packet transmission interval. Upper: TF1, 200 bits/s. Middle: TF2, 400 bits/s. Lower: TF3, 1700 bits/s.

## 4  Sea trials

Several sea trials were conducted during the SensIs project. Here we present the results of the final one, carried out in the Oslo fjord – Horten area, February 10-12 2015. The experiment site has ca 200 m sea depth and approximately flat bottom. A total of 6 deployed subsea nodes (named N2–N7 in the following) and one master node (N1) were used. N1 was mainly on board the vessel Simrad Echo using a dunking transducer at 30 m depth. The configuration was set up to cover approximately 30 % of the full network in Figure 1. The first-day configuration is shown in Figure 3. N2 and N4 were equipped with one Nortek sensor each. Note that these nodes were suspended 100 m above the sea floor. To supplement the real sensors, we had the option of starting simulated sensor traffic from any node by remote control. N6 was used in this way in selected periods.

Earlier SensIs experiments in the same region had demonstrated heavy multipath, dominated by a late-arriving surface reflection. To reduce this interference, specially designed transducers were developed, having a horizontally donut-shaped directivity (-3 dB beamwidth 40º, directivity index (DI) 9dB). These were used for nodes 2, 3 and 7, see Figure 3. The rest of the nodes were using omnidirectional transducers (180º beamwidth, DI 2 dB), except N6 that was fitted with a horizontally oriented cone beam transducer (50º) originally developed for long distance vertical communication.

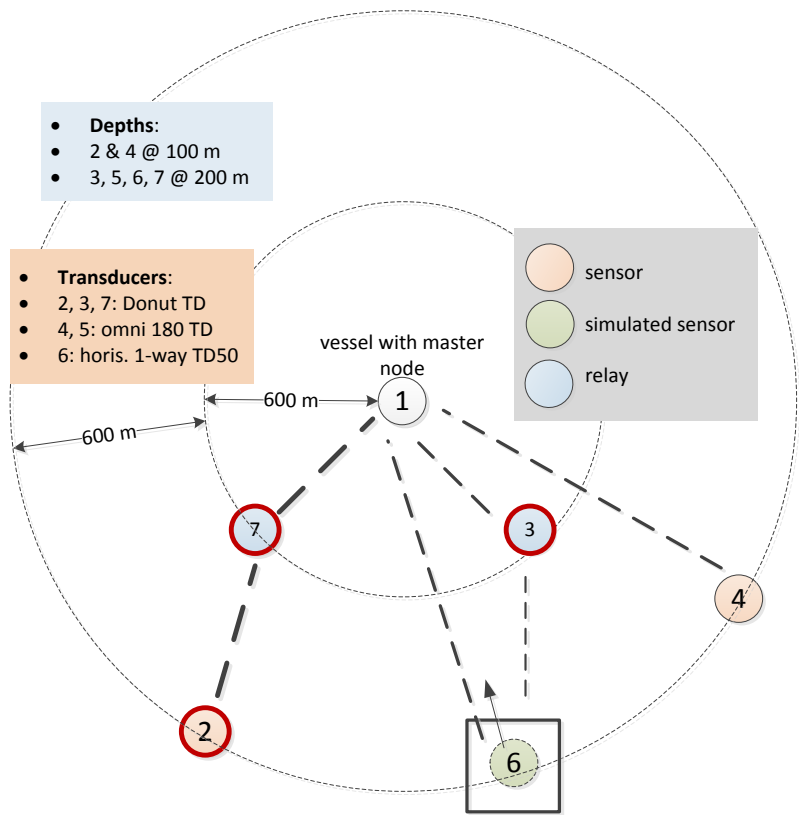

Figure 3 Deployment configuration Day 1, Feb 11. Red circles: donut-directivity transducers. Dashed lines indicate paths selected by the routing protocol.

Sound propagation conditions were assessed by measuring CTDs and using raytracing to estimate impulse responses. The CTDs were virtually constant during the test-period. The results are shown in Figure 4 in the form of eigenrays and estimated SINR vs range. Raytracing was carried out by PlaneRay [7], slightly modified by adding receiver directivity and using [8] for surface bounces and [9] for bottom bounces. The surface bounce model imposes almost zero energy loss at the sea surface, for wide transducer beams and calm seas. Note that surface bounces should be interpreted as mainly scattering rather than specular reflections. The dominating multipath components (surface mainly) always arrive too late to permit any processing gain in the modem. The interference, therefore, is equal for all transport formats. Furthermore, ambient noise is entirely negligible in comparison with the interference.

The main experiments were conducted Feb 11 (Day 1), overnight to Feb 12 (Overnight), and on Feb 12 (Day 2). The configurations were as follows:
- Day 1: Figure 3.
- Overnight test: Upper part of Figure 5. The surface vessel with its sink node was absent, and N7 was reconfigured to be sink, also renaming it N1.

- Day 2: Lower part of Figure 5. The vessel was again present as N1, with the original N7 reset to its Day 1 name and function. Furthermore, the original N3 was replaced by a new hardware, N5 with donut transducer, in order to make sure that its low performance was not due to hardware or software errors. The original N3 was then equipped with an omni-transducer and used as a second relay, close to source node N4.

Note that the main routes chosen by the protocol are shown as dashed lines in Figure 3 and Figure 5.

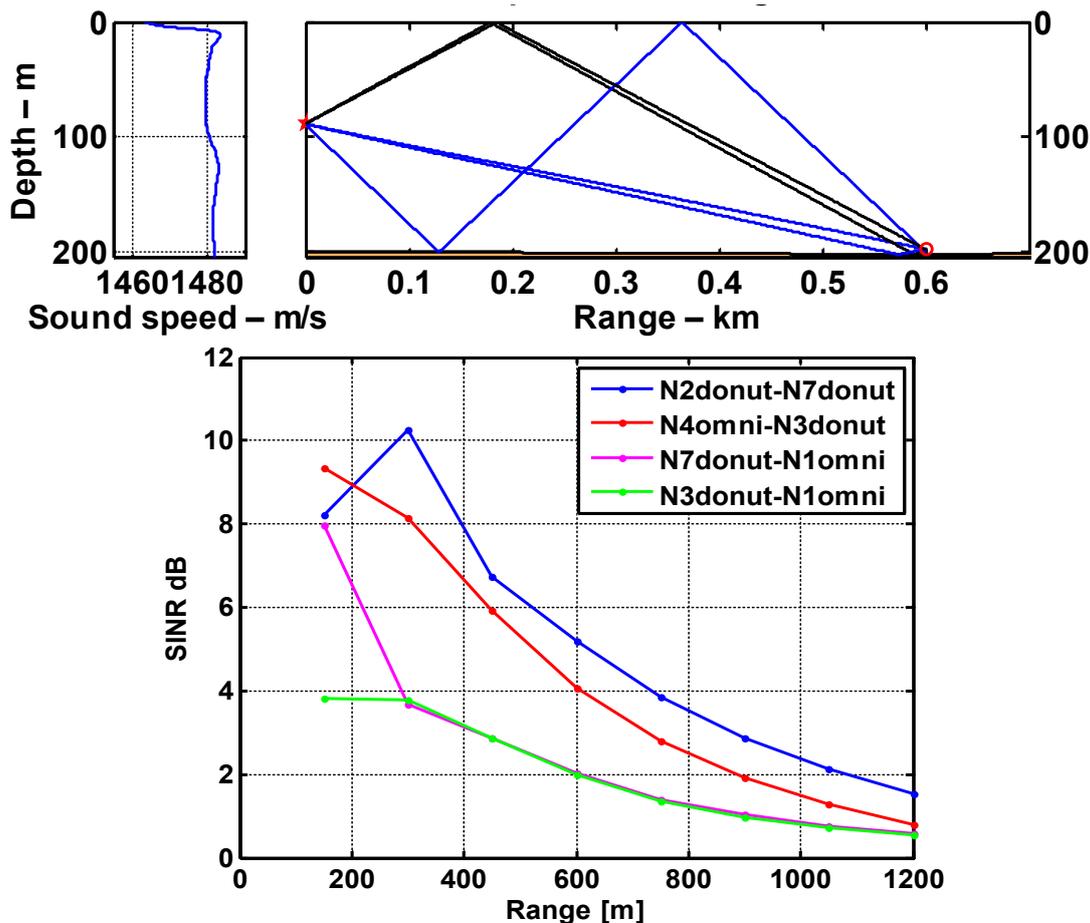

Figure 4 Sound propagation conditions during the experiments. Upper: sound speed profile and an eigenray example (blue and black rays take off downwards and upwards respectively). Lower: SINR estimates vs. range.

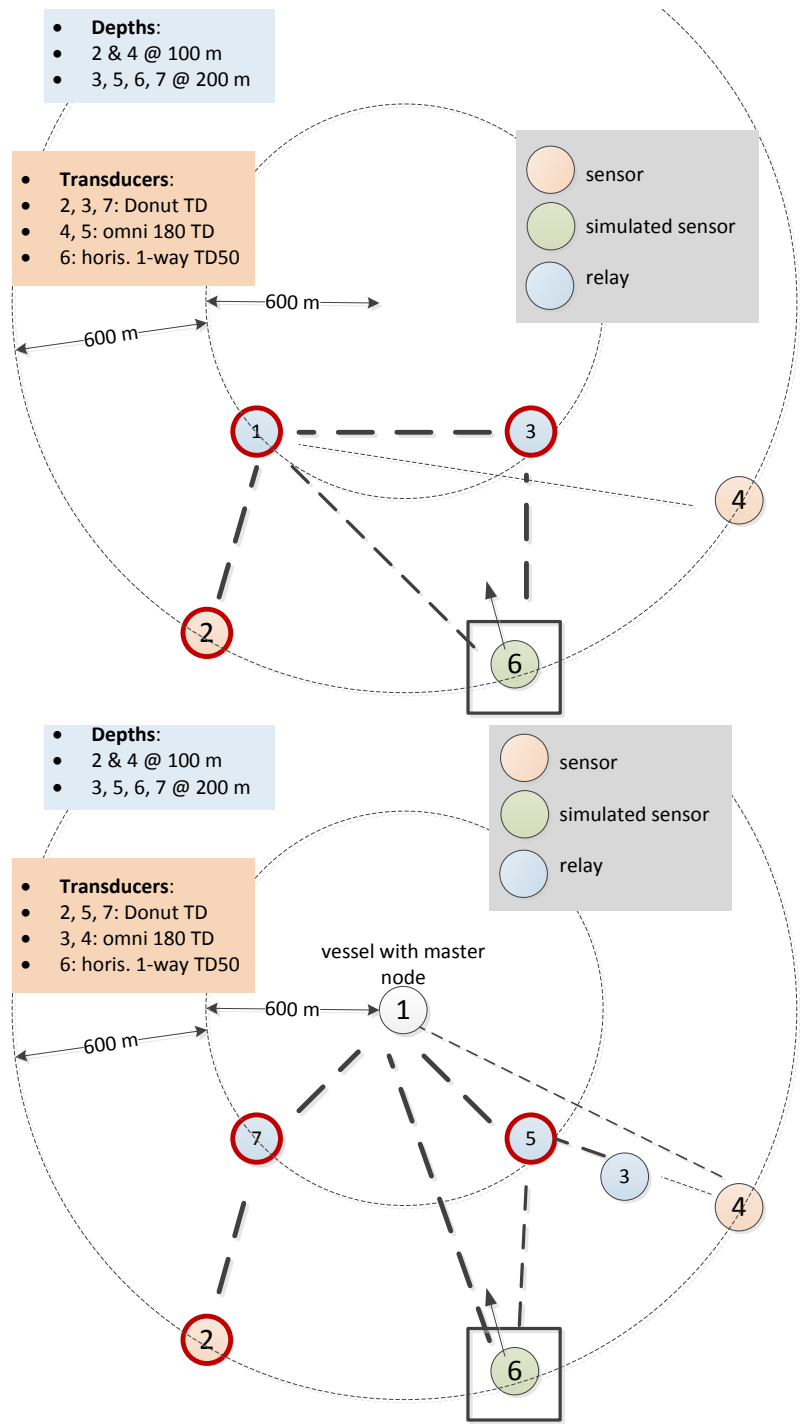

Figure 5 Deployment configurations and node names for the Overnight test (upper) and Day 2 (lower). Dashed lines indicate paths selected by the routing protocol.

The resulting packet delivery ratios for the tests are shown in Table 2. Generally the relay-path N2-N7 was successful, providing high PDRs. Inspection of modem transport formats also revealed that TF3 was used most of the time for the N2-N7 hop and similarly TF2 was chosen for the final hop to N1. Transmission from N4 was, on the other hand, unstable, choosing TF1 much of the time and giving with low PDRs. The main difference between the two source-routes is the use of directive transducers along the N2-route.

Table 2 PDR test results in %

|    | Day 1 | Overnight | Day 2 |
|----|-------|-----------|-------|
| N2 | 75.1  | 95.3      | 46.7  |
| N4 | 22.5  | 2.1       | 9.9   |

Alarm mode was enabled automatically on the two sensor nodes in 15 minute periods, once every whole hour. Due to the more smooth traffic in this mode, the network performance did not change much from normal mode. Simulation results for the full circle setup confirm this, provided that only two sensor nodes at a time are in alarm mode. If alarm mode is activated on all 8 sensor nodes, only a TF3 enabled network would support the needed capacity to avoid drastic performance drop.

## 5 Discussion

The use of directive transducers to reduce the impact of multipath was essential to obtaining good performance in the sea trials. From a protocol perspective this translates into a requirement to have sufficient point-to-point communication capacity to serve the input traffic. In the sea trial results, we see a clear correlation between the ability to use TF2 and TF3 and the ability to obtain stable UC mode periods. Even though TF1 produces reliable link performance, its typical packet duration of 2.1 seconds is just too much to avoid severe network interference due to a very busy water channel. As mentioned in section 4, TF3 is the only one that would be able to serve simultaneous alarms on a full ring of nodes.

The results also showed that ICRP need enough capacity to perform the network discovery and new path search through its use of broadcasting. The ability to run TF2 and TF3 modes was clearly superior in this regard compared to TF1.

Another way to alleviate capacity requirements is to choke the forwarding functionality of BC packets at certain nodes. This would be reasonable for the sensor nodes in our situation, since these form an outer circle and communication fundamentally should flow towards the centre. Simulations at TF1, not detailed in section 3, showed that this would improve the performance but not sufficiently so to obtain satisfactory performance.

We are aware that the selected MAC protocol (CSMA Aloha) has not been a performance-wise optimal choice due to high traffic load imposed by the SensIs scenario, and that some of the performance issues in high load can be coloured by our MAC protocol selection. CSMA Aloha throughput decreases when pushing it towards its capacity limits. CSMA Aloha strengths are flexibility and adaptability. Further work examining ICRP on top of alternative MAC protocols, e.g., such as the distance aware DACAP, could unmask if some of our findings are dependent on the selected medium access protocol.

## 6 Conclusions

The SensIs project has demonstrated that the ICRP provides the necessary functionality and performance to build an underwater acoustic network with a large number of traffic-aggressive sensor nodes. ICRP performance has been investigated in both simulations and real deployments at sea. It provides timely delivery of sensor messages even in congested periods, at the cost of somewhat increased traffic due to more frequent broadcasting periods. All this provided that sufficient point-to-point communication capacity is available.

By providing some new features compared to the original ICRP publication, the protocol could run larger periods in unicast mode, and flooding traffic volume could be relaxed.

By using moderately directive transducers, higher point-to-point reliability and bitrates were obtained, leading to improved packet delivery ratios and route path selections. This measure was found necessary to obtain sufficient performance for the ice monitoring application.

## Acknowledgements

The authors would like to thank the Norwegian Research Council for financial support. We would also like to thank University of Padova for providing the DESERT framework as open software to the underwater communication research community.